\documentclass[pra,twocolumn,amsmath,amssymb,showpacs]{revtex4}
\usepackage{bbm}
\usepackage{amsmath}
\usepackage{amssymb}
\usepackage{mathrsfs}
\usepackage{graphicx}
\begin{document}

\title{Thermal quantum discord in the anisotropic Heisenberg XXZ model with the Dzyaloshinskii-Moriya interaction}%
\author{Yi-Xin Chen}
\email{yxchen@zimp.zju.edu.cn}
\affiliation{Zhejiang Insitute of Modern Physics, Zhejiang University, Hangzhou 310027, China}%
\author{Zhi Yin}
\email{zhiyin@zimp.zju.edu.cn}
\affiliation{Zhejiang Insitute of Modern Physics, Zhejiang University, Hangzhou 310027, China}%
\date{\today}


\begin{abstract}
The thermal quantum discord (QD) is studied in a two-qubit
Heisenberg XXZ system with Dzyaloshinskii-Moriya (DM) interaction.
We compare the thermal QD with thermal entanglement in this system
and find remarkable differences between them. For instance, we show
situations where QD decreases asymptotically to zero with
temperature $T$ while entanglement decreases to zero at the point of
critical temperature, situations where QD decreases with certain
tunable parameters such as $D_z$ and $D_x$ when entanglement
increases. We find that the characteristic of QD is exotic in this
system and this possibly offers a potential solution to enhance
entanglement of a system. We also show that tunable parameter $D_x$
is more efficient than parameter $D_z$ in most regions for
controlling the QD.
\end{abstract}
\maketitle
\section{Introduction}

In recent years, the concept of entanglement \cite{hor09} has
received much attention in quantum information and quantum
computation \cite{nielson} due to its fascinating features in
developing the idea of quantum communication and quantum computer.

Entanglement is one kind of quantum correlations \cite{hor09} and
offers support for lots of quantum tasks such as quantum
teleportation \cite{bennett93,bou97}, quantum dense code
\cite{bennett92} and quantum computing \cite{shor99,gro97}. However,
it is not the only part of quantum correlations \cite{maz09}. So
what is the quantum correlation without entanglement? An answer to
this question is the notion of quantum discord (QD) given by
Ollivier and Zurek \cite{zurek01}. It introduces a measure of the
quantumness of correlations (including entanglement). This notion
has attracted much attention nowadays
\cite{sarandy09,dil08,luo08,wer09,werlang09,maz09,chen09}. It can be
treated as an important quantum resource and even used for the
quantum computation task \cite{dat08,lanyon08,kni98}.

As a kind of quantum resource, entanglement can be generated,
maintained and controlled in so many condensed matter fields
especially solid state systems such as quantum dots
\cite{tra07,hanson07}, Josephson junctions \cite{wei06}, trapped
ions \cite{por04} and cavity-QED systems \cite{zheng00}, etc. Among
systems above, owing to its concise form of Hamiltonian and ability
to be realized in real physical systems, there has been plenty of
work discussing entanglement and thermal entanglement qualitatively
and quantitatively in spin-chain systems
\cite{wang02,sun03,zhang05,arn01,aso05,gun01}.

Since much work discuss the spin-spin interaction in spin-chain
systems such as Heisenberg model, more and more researches study the
influence of Dzyaloshinskii-Moriya (DM) interaction
\cite{dzy58,mo60} which is arising from spin-orbit interaction in
spin-chain systems \cite{aristov00,kargarian09,jafari08}. Recently,
there is an interesting paper \cite{li08} investigating the thermal
entanglement in an anisotropic XXZ model with DM interaction. The
characteristic of thermal entanglement in that model is notable
especially when you control the entanglement via tuning
$x$-component DM interaction.

Since QD is considered as quantum correlation and quantum resource
as well as entanglement. It is curious to ask what the
characteristic of QD is at finite temperature, and what the
differences are between thermal QD and entanglement in this system.
In our paper we introduce this two-qubit anisotropic Heisenberg XXZ
model with DM interation. We discuss the dependence of thermal QD in
this system on temperature $T$ and some other tunable parameters
such as DM interaction parameters $D_z$, $D_x$. We also compare the
QD with entanglement measured by concurrence
\cite{1bennett96,2bennett96,wootters98} between two qubits in the
model and find remarkable differences between them. QD is more
robust than entanglement concurrence versus temperature $T$. (This
property can also be found in Ref. \cite{werlang09} where a
Heisenberg XYZ model with a magnetic term was considered.) We find
that the characteristic of QD is special in this system since QD
decreases with certain tunable parameters such as $D_z$ and $D_x$
when entanglement increases. And this possibly offers a potential
solution to enhance entanglement of a system. Furthermore, we
analyze the influence of different DM interaction parameters $D_z$
and $D_x$ on QD and reveal that tunable parameter $D_x$ is more
efficient than parameter $D_z$ in most regions for controlling the
QD.

Our paper is organized as follows. In Sec. II we review the concept
of quantum discord. In Sec. III we investigate the thermal quantum
discord in two XXZ Heisenberg model with different DM interaction
parameters. We show the main results and compare thermal QD with
thermal entanglement. In Sec. IV we discuss the opposite tendency of
thermal QD and entanglement. Conclusion are then presented in Sec.
V.

\section{Quantum Discord}

In classical information theory, the correlation between two random
variables $A$ and $B$ are described by mutual information which
reads in two expressions.
\begin{equation}
\mathcal{I}(A:B) = H(A)+H(B)-H(A,B) \label{i1}\end{equation}
\begin{equation}
\mathcal{J}(A:B) = H(A)-H(A|B) \label{i2}\end{equation} Here,
$H(A)=-\sum_{a}p_{a}\log_{2}p_{a}$ is Shannon entropy. $H(A,B)$ is
the joint Shannon entropy which is defined by
$H(A,B)=-\sum_{a,b}p_{a,b}\log_{2}p_{a,b}$. $H(A|B)$ is conditional
entropy and introduced for quantifying the ignorance (on average)
about the value of A given B is known. This two expressions are
equivalent using Bayes' rule:
$p(a_{i},b_{j})=p(a_{i}|b_{j})p(b_{j})=p(b_{j}|a_{i})p(a_{i})$,
where $p(a_{i}|b_{j})$ is the conditional probability.

In order to generalize these expressions above to quantum case, we
replace classical probability distributions by density matrices and
the Shannon entropy by von Neumann entropy
$S(\rho)=-\textrm{tr}(\rho\log\rho)$. Since we can denote the
density matrix of a composite system $AB$ by $\rho^{AB}$ and the
density matrices of parts $A$ and $B$ by $\rho^A$ and $\rho^B$,
respectively, we can rewrite Eq. (\ref{i1}) and Eq. (\ref{i2}) as
\begin{equation}
\mathcal{I}(\rho^A:\rho^B) = S(\rho^A)+S(\rho^B)-S(\rho^{AB})
\label{q1}\end{equation}
\begin{equation}
\mathcal{J}(\rho^A:\rho^B) = S(\rho^A)-S(\rho^A|\rho^B),
\label{q2}\end{equation} where $S(\rho^{AB})$ is quantum joint
entropy and $S(\rho^A|\rho^B)$ is quantum condition entropy.
However, here we can not just obtain quantum condition entropy
simply via replacing Shannon entropy by von Neumann entropy. Due to
the definition of condition entropy, $S(\rho^A|\rho^B)$ depends on
the choice of measurement. To get quantum condition entropy, we
should choose a measurement which is described by a set of
projectors $\{\hat{\Pi}_{i}^{B}\}$ performed locally on B. After the
measurement, the state of system is
$\rho_{i}=\hat{\Pi}_{i}^{B}\rho^{AB}\hat{\Pi}_{i}^{B}/p_{i}$ where
$p_{i}=\textrm{tr}(\hat{\Pi}_{i}^{B}\rho^{AB}\hat{\Pi}_{i}^{B})$.
Now we can get the definition of quantum condition entropy.
Following the Eq. (\ref{q2}), another expression of quantum mutual
information is
\begin{equation}
\mathcal{J}(\rho^{AB}:\{\hat{\Pi}_{i}^{B}\}) =
S(\rho^{A})-S(\rho^{AB}|\{\hat{\Pi}_{i}^{B}\}) \label{mut}
\end{equation}
Since different choices of $\{\hat{\Pi}_{i}^{B}\}$ decide different
$\mathcal{J}(\rho^{AB}:\{\hat{\Pi}_{i}^{B}\})$, Eq. (\ref{q1}) and
Eq. (\ref{q2}) may not be equal any more. According to the original
definition \cite{zurek01}, quantum discord is the minimum of
difference between Eq. (\ref{q1}) and Eq. (\ref{mut}), i.e.
\begin{equation}
D(\rho^{AB})=\textrm{min}\left[\mathcal{I}(\rho^A:\rho^B)-\mathcal{J}(\rho^{AB}:\{\hat{\Pi}_{i}^{B}\})\right]\end{equation}
It is a measure of the quantumness of a pairwise correlation. With
the expression of quantum mutual information of Eq. (\ref{mut}),
following Ref. \cite{zurek01,henderson01}, we can define classical
correlation between $A$ and $B$ as
\begin{equation}
CC(\rho^{AB})=\underset{{\{\hat{\Pi}_{i}^{B}\}}}{max}(\mathcal{J}(\rho^{AB}:\{\hat{\Pi}_{i}^{B}\}))\label{classical}
\end{equation}
Eventually, we get quantum discord in another expression:

\begin{equation}
Q(\rho^{AB})=\mathcal{I}(\rho^A:\rho^B)-CC(\rho^{AB})\label{discord}
\end{equation} Here the quantum mutual information is used as a measure of total
correlations \cite{verdral02,gro05}.

Quantum discord is a measure to quantify all the quantum
correlations including entanglement of a pairwise correlation. For
two subsystems $A$ and $B$ are correlated classically only, the QD
is zero. Moreover, for some pairwise system such as Werner states
\cite{zurek01}, QD is non-zero although the system is separable. In
this sense, some two-body states that don't have entanglement do not
mean there is no quantum correlations between them. We can even use
this kind of quantum correlation power to do some quantum
computation tasks \cite{dat08,lanyon08,kni98}.

\section{XXZ Heisenberg model with different DM interaction parameters}

In this section, we introduce an anisotropic XXZ Heisenberg model
with different DM interaction parameters. We show the dependence of
thermal QD on temperature $T$ and some other tunable parameters such
as $J_z$, $D_z$ and $D_x$ in this model. We compare the thermal QD
with thermal entanglement concurrence in the same model and find
some attractive results.

\subsection{XXZ Heisenberg model with DM interaction parameter $D_z$}

The Hamiltonian $H$ of a two-qubit anisotropic Heisenberg XXZ chain
with DM interaction parameter $D_z$ is
\begin{equation}
H=J\sigma_1^x\sigma_2^x+J\sigma_1^y\sigma_2^y+J_z\sigma_1^z\sigma_2^z+D_z(\sigma_1^x\sigma_2^y-\sigma_1^y\sigma_2^x).
\label{h1}\end{equation} where $J$ and $J_z$ are the real coupling
coefficients, $D_z$ is the $z$-component parameter of the DM
interaction. Here the so-called DM interaction is a supplemented
magnetic term arising from the spin-orbit coupling
\cite{dzy58,mo60}. $\sigma^i$ ($i = x, y, z$) are Pauli matrices. In
our paper, we consider the antiferromagnetic case ($J>0$, $J_z>0$).
This model is reduced to the isotropic XX model when $J_z=0$ and to
the isotropic XXX model when $J_z=J$. Parameters $J$, $J_z$ and
$D_z$ are dimensionless here.

First we show the matrix form of this $H$ in the standard basis
$\{|00\rangle,|01\rangle,|10\rangle,|11\rangle\}$, and get
\begin{equation}
H = \left(
\begin{array}{cccc}
J_z&0&0&0\\
0&-J_z&2J+2iD_z&0\\
0&2J-2iD_z&-J_z&0\\
0&0&0&J_z
\end{array}
\right)\label{h1m}
\end{equation}

Now we want to consider the state of the system as thermal
equilibrium state in a canonical ensemble. The state of spin chain
system at thermal equilibrium is given by the Gibb's density
operator $\rho(T)=\frac{exp(-\beta H)}{Z}$ where $Z=tr[\exp(-\beta
H)]$ is the partition function of the system, $H$ is the system
Hamiltonian and $\beta=\frac{1}{K_BT}$ with T temperature. $K_B$ is
the Boltzmann's constant which we take equal to 1 for simplicity.
Here $\rho(T)$ presents a thermal state. Plenty of paper have
discussed about entanglement in thermal state
\cite{wang02,sun03,zhang05,arn01,aso05,gun01} which is called
thermal entanglement \cite{nielsen00}. In our paper, we discuss
another kind of quantum correlation-quantum discord at finite
temperature i.e., thermal quantum discord.

By some routine calculations, we can get expression of $\rho(T)$:
\begin{equation}
\rho(T) =\frac{1}{Z} \left(
\begin{array}{cccc}
e^{-\beta J_z}&0&0&0\\
0&u&ve^{i\theta}&0\\
0&ve^{-i\theta}&u&0\\
0&0&0&e^{-\beta J_z}
\end{array}
\right)
\end{equation}
where $u=\frac{1}{2}(1+e^{4\beta w})e^{\beta(J_z-2w)}$,
$v=\frac{1}{2}(1-e^{4\beta w})e^{\beta(J_z-2w)}$, $Z=2e^{-\beta
J_z}[1+e^{2\beta J_z}\cosh(2\beta w)]$ and $w=\sqrt{J^2+D_z^2}$,
$\theta=\arctan(\frac{D_z}{J})$.

According to Eq. (\ref{discord}), to obtain QD of a two-qubit
system, we should first get $\rho^A$, $\rho^B$ for subsystems $A$
and $B$ respectively and $\rho^{AB}$ for the composite system. Since
we have $\rho^{AB}$ already, we need to evaluate $\rho^A$, $\rho^B$.
Choosing $|0\rangle$, $|1\rangle$ as a basis, we can trace out the
rest of the subsystem to get reduced density matrix of $A$ and $B$
employing formulas below.
\begin{equation}
\rho^A\equiv tr_B\rho^{AB}\equiv
\sum_{j=1}^{N_B}(I_A\bigotimes\langle\phi_j|)\rho^{AB}(I_A\bigotimes|\phi_j\rangle)
\end{equation}

\begin{equation}
\rho^B\equiv tr_A\rho^{AB}\equiv
\sum_{j=1}^{N_A}(\langle\psi_j|\bigotimes
I_B)\rho^{AB}(|\psi_j\rangle\bigotimes I_B)
\end{equation} where $I_A$ and $I_B$ are the identity operators, $|\phi_j\rangle$ ($j=1, 2, ..., N_B$) and $|\psi_j\rangle$ ($j=1, 2,
..., N_A$) are orthonormal bases in $\mathcal{H}_B$ and
$\mathcal{H}_A$ respectively.

Now we can get $\rho_A=\rho_B=\frac{1}{Z}(e^{-\beta J_z}+u)I$ where
$I$ is the identity operator. Next we evaluate quantum discord
numerically according to the following instruction. Choose the set
of projectors
$\{|\psi_1\rangle\langle\psi_1|,|\psi_2\rangle\langle\psi_2|\}$,
where
$|\psi_1\rangle=\cos{\theta}|0\rangle+e^{i\phi}\sin{\theta|1\rangle}$
and
$|\psi_2\rangle=-\cos{\theta}|1\rangle+e^{-i\phi}\sin{\theta|0\rangle}$
, to measure one of the subsystems. The classical correlation
$CC(\rho^{AB})$ is obtained numerically by varying the angles
$\theta$ and $\phi$ from $0$ to $2\pi$ and eventually according to
Eq. (\ref{discord}) we can get QD as well.

\begin{figure}
[ptbh](a)
    \begin{minipage}{2.4in}
        \includegraphics[
         width=2.4in,height=1.4in 
        ]
            {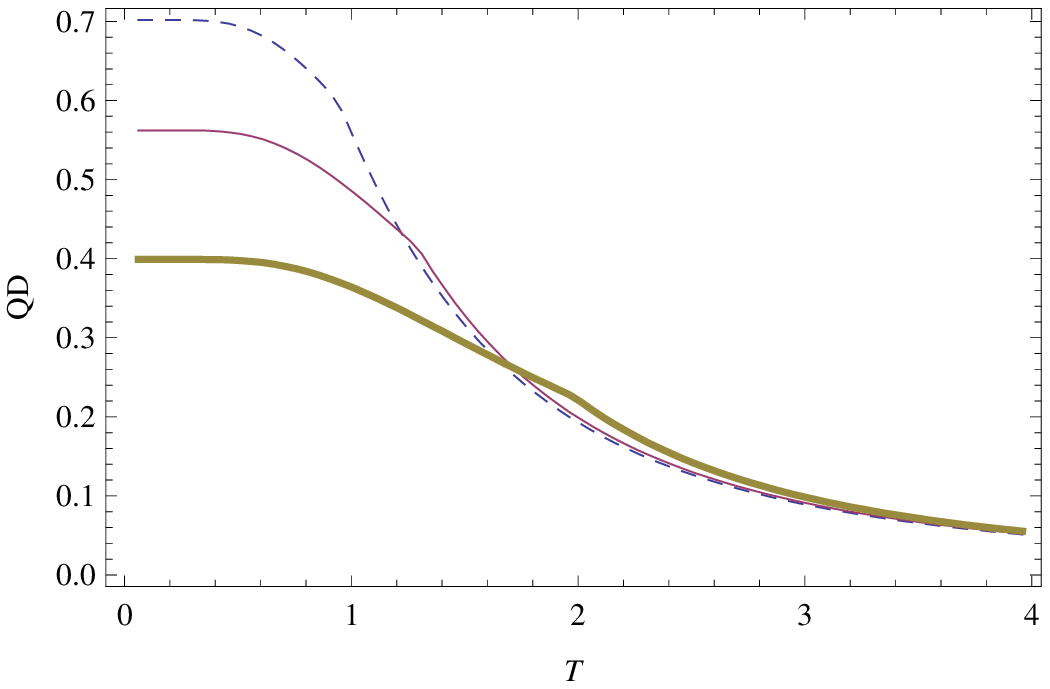}
    \end{minipage}
    \hspace{0.0in}

    (b)
    \begin{minipage}{2.4in}
            \includegraphics[
             width=2.4in,height=1.4in 
            ]
            {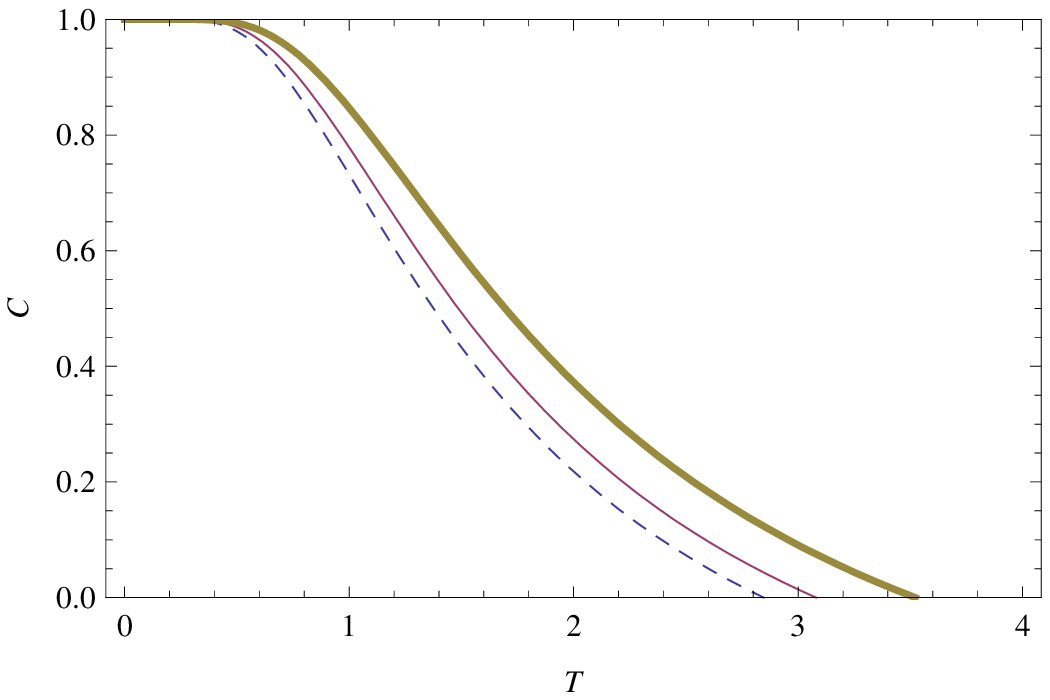}%
    \end{minipage}
    \hspace{0.0in}

    (c)
    \begin{minipage}{2.4in}
            \includegraphics[
             width=2.4in,height=1.4in 
            ]
            {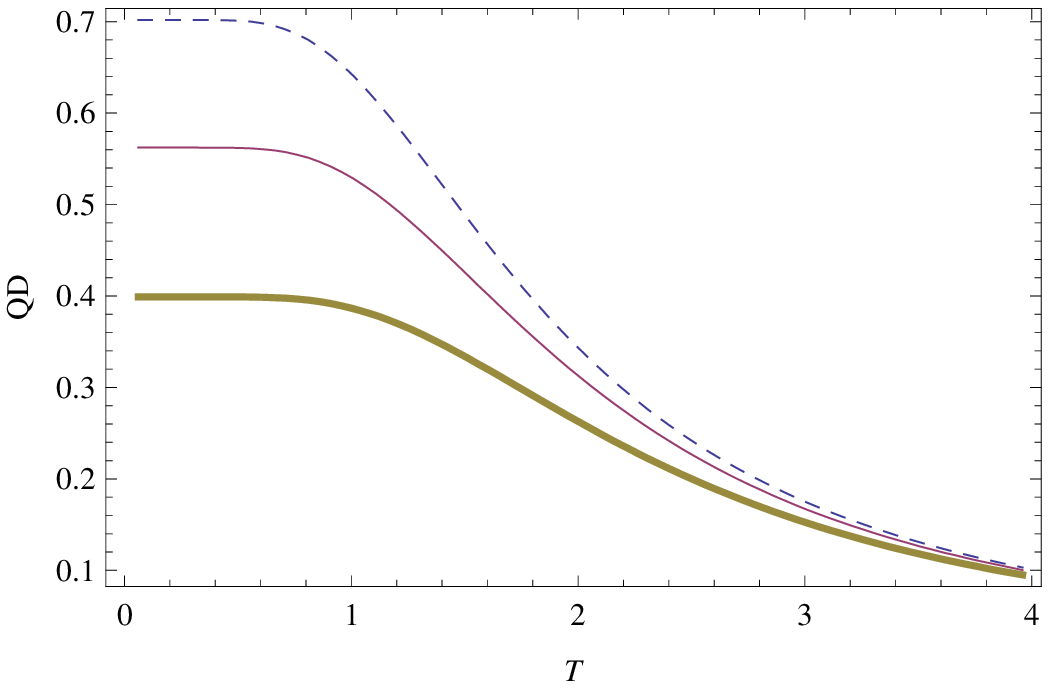}%
    \end{minipage}

    \caption{The QD (a), (c) and concurrence (b) versus T in the system with $z$-component parameter $D_z$. Here $J=1$.
    (a) Here $J_z=0.2$ and for the thick line $D_z=1.0$, normal line $D_z=0.7$, dashed line $D_z=0.5$.
    (b) Here $J_z=0.2$ and for the thick line $D_z=1.0$, normal line $D_z=0.7$, dashed line $D_z=0.5$.
    (c) Here $J_z=1$ and for the thick line $D_z=1.0$, normal line $D_z=0.7$, dashed line $D_z=0.5$.}

\end{figure}

Using the method in Ref. \cite{1bennett96,wootters98,2bennett96},
and results in \cite{li08}, we can also simply get the concurrence
(In this paper, we use concurrence as the measure of entanglement)
of this model at finite temperature. $C(\rho(T))=\frac{\beta
J_z}{Z}(e^{2\beta w}-e^{-2\beta J_z}-e^{-2\beta w}-e^{-2\beta J_z})$
if $J_z>-w$, and $C(\rho(T))=0$ if $J_z<-w$.

Now by fixing some parameters we can see the variations of thermal
QD and entanglement under the influence of other parameters. Looking
at Fig. 1, where $J=1$, we can find that the QD decreases with
increasing of temperature $T$. This tendency is similar to the
situation of entanglement as shown in panel (b) where the conditions
are the same. Differently, we can see from panel (b) that thermal
entanglement will vanish when $T$ reaches certain point. And the
larger parameter $D_z$ is the larger critical point is. The point
denotes so-called critical temperature of the system. This behavior
is caused by mixing of the maximally entanglement state with other
states. In other words, it means that with the increasing of
temperature $T$, thermal entanglement of the system will be
disappeared at critical temperature and there is no entanglement
resource to use then. However, as we see in panels (a) and (c),
thermal QD will not reach zero point. It acts asymptotically near
zero when the temperature is high. And in the temperature regions
where thermal entanglement is zero while QD is still alive. We even
find that we adjust $T$ very high (not shown in the plots), QD
becomes very small but nonzero. This interesting phenomenon can be
explained as follows. With the temperature grows, the role of
thermal fluctuations exceeds quantum ones. However, it cannot kill
the quantumness. As we know, entanglement is not the only part of
quantum correlations. Based on Ollivier and Zurek's argument
\cite{zurek01}, it shows that absence of entanglement does not imply
classcality. If you want to destroy the quantumness of a system, you
should use the process of decoherence. Decoherence will eventually
make the quantumness of a system to be zero and then the system will
be in totally classical state. Since QD measures total quantum
correlations, it is obvious that even temperature is high, without
introducing the decoherence, quantum correlations will not vanish.
So in this sense, QD is robust than entanglement at a finite
temperature.

Focusing now at panel (a) of Fig. 1, when temperature $T$ increases,
thermal fluctuation takes control of the system in this region, and
the role of coupling constant such as $J$, $J_z$ and $D_z$ become
weaker. So we can see the lines in panel (a) of Fig. 1 become
concentrated together. Now we enhance the parameter $J_z$ to 1.0 as
we can see in panel (c) of Fig. 1. We find that at the same
temperature, lines become more separate than the case of panel (a).
This effect is because of the influence of coupling constant $J_z$
makes the quantumness considerable. Next, let us take a look at
panels (a) and (c) again and consider the influence of parameter
$D_z$ to thermal QD. Due to the symmetry of DM interaction
parameter, we can only take the strength of $D_z$ into account. So
here only consider the $D_z>0$ case. Apparently, we can see a
remarkable phenomenon (Also shown in Fig. 3) in certain region that
when the parameter $D_z$ increases, thermal QD decreases while this
trend of QD is in contrast to the behavior of entanglement. Since QD
and entanglement are both kinds of measure of quantum correlations.
This is very wired. Worth to be mentioned, this phenomenon appears
in $D_x$ case as well, and we will analyze this in detail in Sec.
IV.

Moreover, take a look at Fig. 2, where we fix $J=1$ and $D_z=1$.
Asymptotical phenomenon of QD also appears at finite temperature for
different $J_z$ in panel (a). However, we can see that QD increases
with increasing of $J_z$. This is corresponding to the case of
entanglement shown in panel (b). Here, $J_z$ as a anisotropic
coupling constant of Heisenberg model, plays the same role in
thermal QD as in the case of entanglement.

\begin{figure}
[ptbh](a)
    \begin{minipage}{2.4in}
        \includegraphics[
         width=2.4in,height=1.4in 
        ]
            {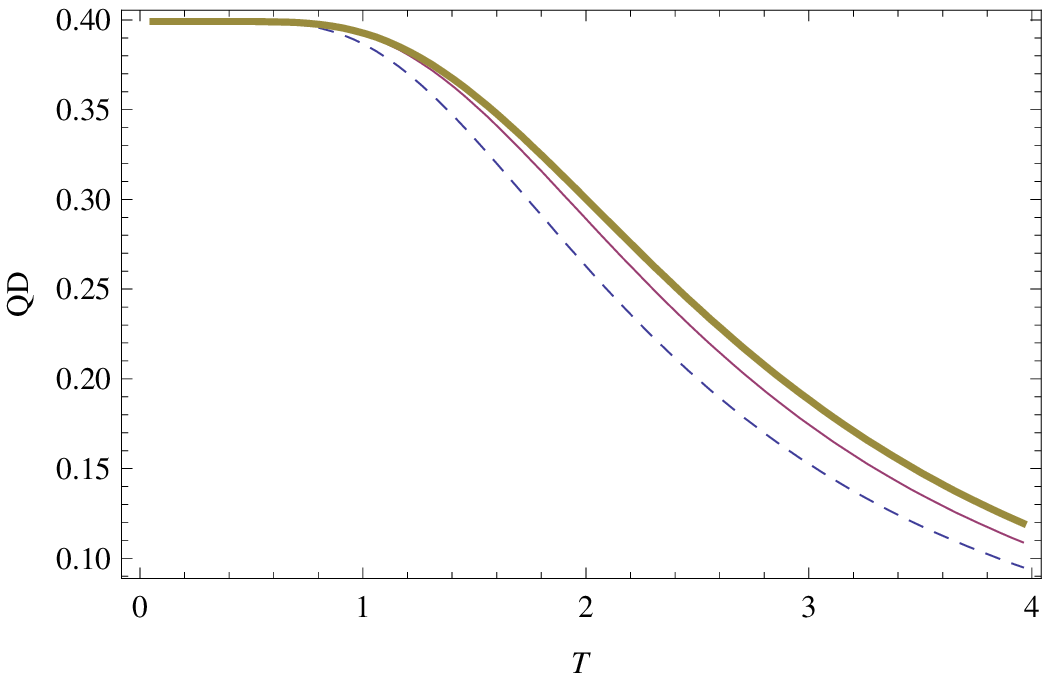}
    \end{minipage}
    \hspace{0.0in}

    (b)
    \begin{minipage}{2.4in}
            \includegraphics[
             width=2.4in,height=1.4in 
            ]
            {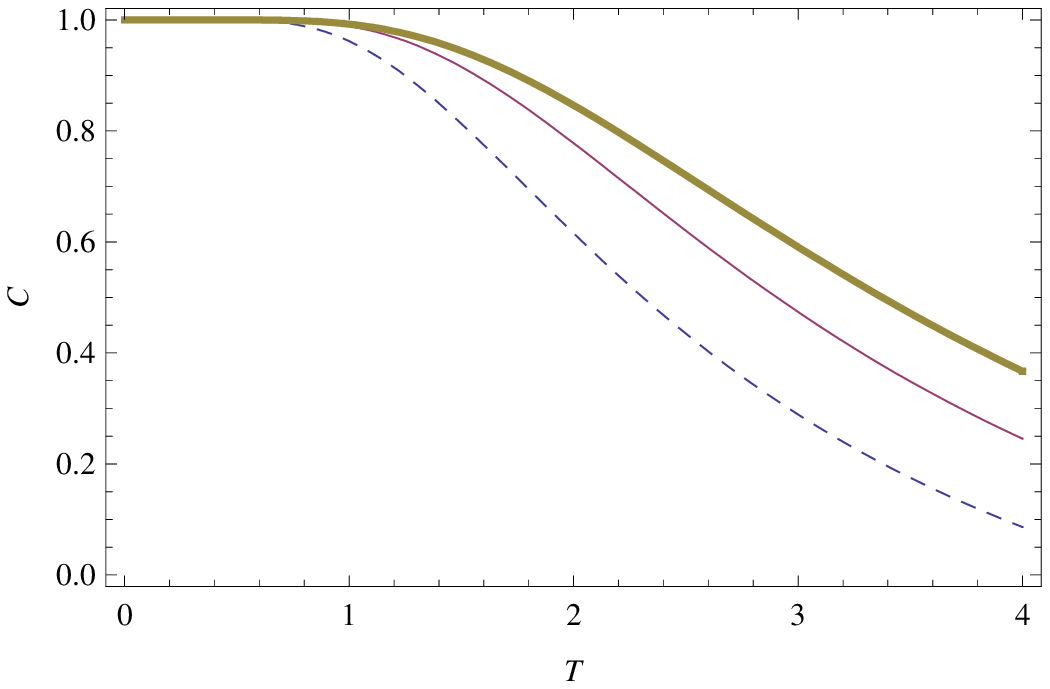}%
    \end{minipage}

    \caption{The QD (a) and concurrence (b) versus T in the system with $z$-component parameter $D_z$. Here $J=1$, $D_z=1$.
    (a) Here for the thick line $J_z=3.0$, normal line $J_z=2.0$ and dashed line $J_z=1.0$.
    (b) Here for the thick line $J_z=3.0$, normal line $J_z=2.0$ and dashed line $J_z=1.0$.}

\end{figure}

\begin{figure}
\includegraphics[width=7cm]{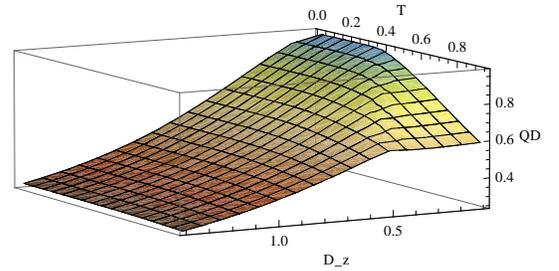}

\caption{(Color online) The quantum discord is plotted versus $T$
and $D_z$ where $J=1$ and $J_z=0.2$.}

\end{figure}

\subsection{XXZ Heisenberg model with DM interaction parameter $D_x$}

Here we consider the case of the two-qubit anisotropic Heisenberg
XXZ chain with DM interaction parameter $D_x$. The Hamiltonian $H'$
of this model reads

\begin{equation}
H'=J\sigma_1^x\sigma_2^x+J\sigma_1^y\sigma_2^y+J_z\sigma_1^z\sigma_2^z+D_x(\sigma_1^y\sigma_2^z-\sigma_1^z\sigma_2^y).
\end{equation}
where $D_x$ is the $x$-component parameter of the DM interaction and
other parameters are the same as in part A of this section.

Following the exact same approach in part A, we can get

\begin{equation}
H' = \left(
\begin{array}{cccc}
J_z&iD_x&-iD_x&0\\
-iD_x&-J_z&2J&iD_x\\
iD_x&2J&-J_z&-iD_x\\
0&-iD_x&iD_x&J_z
\end{array}
\right)
\end{equation}

\begin{equation}
\rho'(T) =\frac{1}{2Z'} \left(
\begin{array}{cccc}
\mu_+ &-\xi &\xi &\mu_-\\
\xi &\nu_+ &\nu_- &-\xi\\
-\xi &\nu_- &\nu_+ &\xi\\
\mu_- &\xi &-\xi &\mu_+
\end{array}
\right)
\end{equation}
where $w'=\sqrt{(J+J_z)^2+4D_x^2}$
$\phi=\arctan(\frac{2D_x}{J+J_z-w'})$
$\varphi=\arctan(\frac{2D_x}{J+J_z+w'})$, and

$\mu_\pm=e^{-\beta
J_z}\pm(e^{\beta(J-w')}\sin^2{\phi}+e^{\beta(J+w')}\sin^2{\varphi})$

$\nu_\pm=e^{\beta
(J_z-2J)}\pm(e^{\beta(J-w')}\cos^2{\phi}+e^{\beta(J+w')}\cos^2{\varphi})$

$\xi=ie^{\beta(J-w')}\sin{\phi}\cos{\phi}+ie^{\beta(J+w')}\sin{\varphi}\cos{\varphi}$

$Z'=2e^{-\beta J}\cosh[\beta(J-J_z)]+2e^{\beta J}\cosh(\beta w')$.

Also we can get $\rho'^A$ and $\rho'^B$ via straightforward
calculation: $\rho'_A=\rho'_B=\frac{1}{2Z'}(\mu_++\nu_+)I$ where $I$
is the identity operator.

Now, first we look at Fig. 4, where $J=1$ and $D_x=1$. We can see
that when $J_z$ increases, the thermal QD becomes higher. This trend
of QD is the same like entanglement case. (For better comparison we
choose the exact the same value of parameters in Fig. 4 as in Ref.
\cite{li08}) And also, this fits with the behavior of $J_z$ in the
system with $z$-component parameter $D_z$.

While looking at Fig. 5, we find that besides the asymptotical
phenomenon mentioned above, we can also find the opposite trend of
thermal QD and entanglement. So here we can derive that different DM
parameters can both influence the behavior of QD and entanglement
and make them move towards different directions in certain region.
We will address this in Sec. IV.

In addition, back to the Fig. 1 and Fig. 5 which demonstrate the
behaviors of thermal QD in aniosotropic Heisenberg model with
parameters $D_z$ and $D_x$. It is obvious there are some differences
between two different cases. First, let us look at panel (a) of Fig.
1 and panel (b) of Fig. 5, we find that in the situation where other
parameters are exactly the same, $D_x$ is more reliable than $D_z$
when T becomes higher. This indicates that $D_z$ is easy to
influence by the increasing of $T$ (This also shown in Fig. 3 and
Fig. 6). And from Fig. 1 and Fig. 5 we can find that if we tune the
parameter $J_z$ larger, thermal QD depending on different $D_x$
change more obviously than the case of parameter $D_z$. This means
that parameter $D_x$ is more sensitive in certain condition along
with change of parameters such as $J_z$. So we can derive from this
results that it is more efficient to use parameter $D_x$ to control
the QD. In addition, we can also find that when tune parameter $J_z$
to 1.0, the two plots (panel (c) of Fig. 1 and panel (a) of Fig. 5)
become similar, which is mainly because the DM interactions now are
not the dominating parameters controlling the behavior of the
system.

\begin{figure}
\includegraphics[width=7cm]{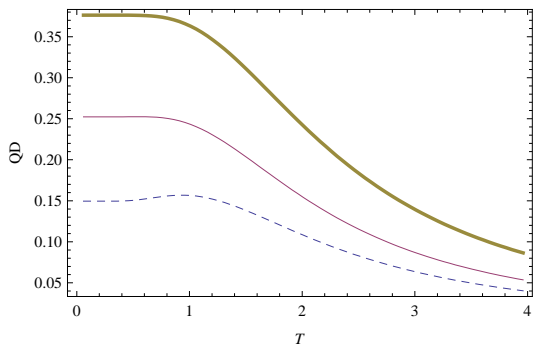}

\caption{The QD versus T in the system with $x$-component parameter
$D_x$. Here $J=1$, $D_x=1$ and for the thick line $J_z=0.9$, normal
line $J_z=0.4$, dashed line $J_z=0$.}

\end{figure}


\begin{figure}
[ptbh](a)
    \begin{minipage}{2.4in}
        \includegraphics[
         width=2.4in,height=1.4in 
        ]
            {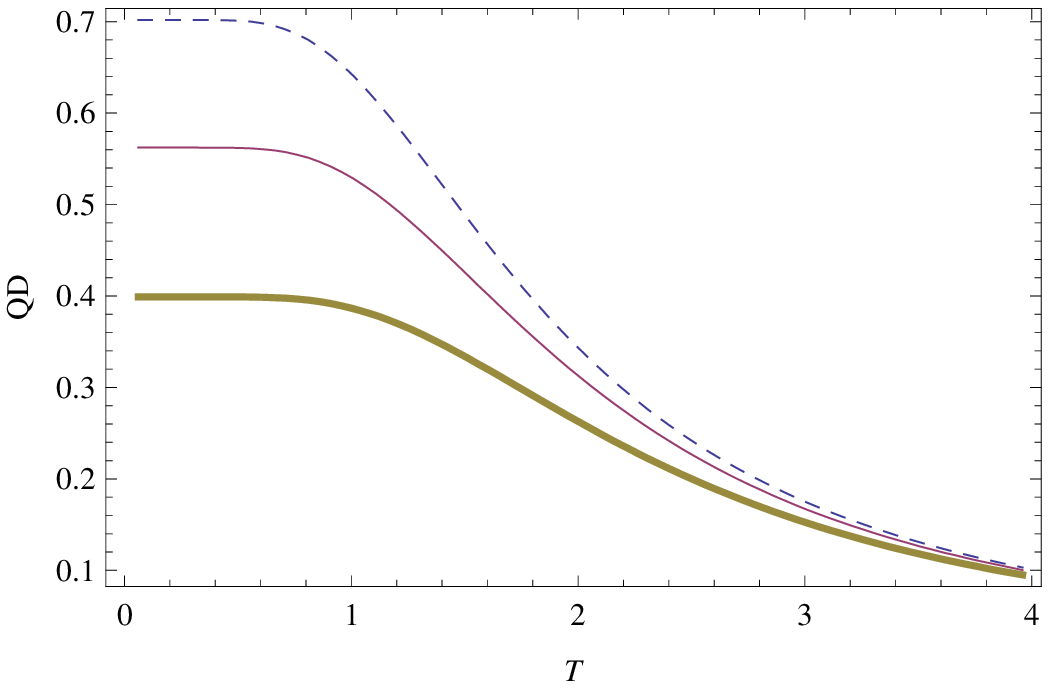}
    \end{minipage}
    \hspace{0.0in}

    (b)
    \begin{minipage}{2.4in}
            \includegraphics[
             width=2.4in,height=1.4in 
            ]
            {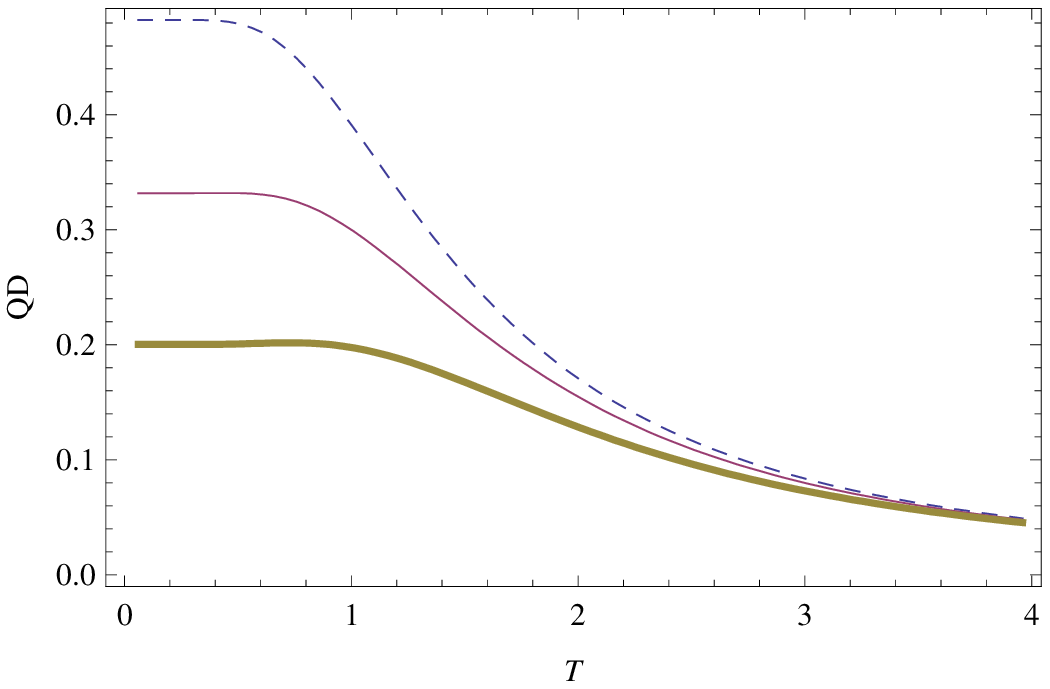}%
    \end{minipage}

    \caption{The QD versus T in the system with $x$-component parameter $D_x$. Here $J=1$.
    (a) Here $J_z=1.0$ and for the thick line $D_z=1.0$, normal line $D_z=0.7$, dashed line $D_z=0.5$.
    (b) Here $J_z=0.2$ and for the thick line $D_z=1.0$, normal line $D_z=0.7$, dashed line $D_z=0.5$. }

\end{figure}


\begin{figure}
\includegraphics[width=7cm]{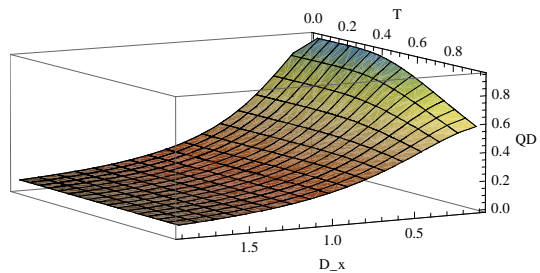}

\caption{(Color online) The quantum discord is plotted versus $T$
and $D_x$ where $J=1$ and $J_z=0.2$.}

\end{figure}

\section{Discussion about opposite tendency of thermal QD and entanglement}
In this section, we will discuss the opposite tendency of thermal QD
and entanglement appeared in Sec. III.

In previous section, we find that, in certain regions, when
parameter $D_z$ increases, thermal QD decreases while at the same
time thermal entanglement increases. This also appears in the case
of parameter $D_x$. So we say that when we tune the parameters of DM
interaction, entanglement and QD have the opposite trend of
movement. Obviously, this notable phenomenon is very wired since
both QD and entanglement quantify the quantumness of correlations.
There are some underlying reasons about that. Although QD and
entanglement both kinds of measure of quantum correlations.
Entanglement is a non-local correlation while QD figures a measure
of whole quantumness of correlations for a pairwise system. There
are some other quantum correlations without entanglement which can
be characterized by QD but not any kinds of entanglement measure
such as concurrence. And we reveal that these other quantum
correlations have remarkably different characteristic from the
entanglement. Recent researches \cite{dat08,lanyon08,kni98} also
suggest that these other quantum correlations can also be an
important resource and maybe take respondence in mixed state quantum
computation. Secondly, DM interaction is a magnetic term arising
from the spin-orbit coupling. The staggered DM interactions bring on
helical magnetic structures due to its form
$\mathbf{D}\cdot[\mathbf{S_1}\times\mathbf{S_2}]$, where
$\mathbf{S_1}$, $\mathbf{S_2}$ denote the spins and $\mathbf{D}$ is
DM vector. DM interaction can contribute the strong planar quantum
fluctuations which pose the alignment ordering to both qubit-1 and
qubit-2 in two-qubit system. Some researches
\cite{jafari08,kargarian09} have shown that in certain condition, DM
interaction can build entanglement. So it is acceptable that
entanglement increase with increasing of DM interaction in this
paper.

Now we can see that since QD figures the whole quantum correlations
including entanglement, the phenomenon here indicate that, the other
quantum correlations without entanglement we mentioned above
decrease rapidly (more rapidly than the increase of entanglement at
the same time) with increasing of parameters of DM interaction. So
we can see from Fig. 1 and Fig. 5. that when QD decreases while the
entanglement increases with the increasing of DM interactions. Here,
this phenomenon can coincide what we have raised above that the
characteristic of QD is exotic. Furthermore, using this
characteristic, under certain conditions, we can tune the parameters
of DM interactions to increase the entanglement of a system and
meanwhile decrease the other quantum correlations of the same
system. The advantage of the method is obvious. If you want to take
entanglement as a quantum resource to do quantum tasks, you want to
enhance your entanglement resource and make the other quantum
correlations of system as less as possible, this approach would
offer a possible solution.

\section{Conclusion}

The thermal QD of a two-qubit anisotropic Heisenberg XXZ chain with
two distinct DM interaction parameters is investigated. We find that
thermal QD is more robust than thermal entanglement versus
temperature $T$. And numerical results indicate that it is more
efficient to control QD in the model with parameter $D_x$ than with
parameter $D_z$. Furthermore, with the increasing of parameters
$D_z$ and $D_x$, thermal QD varies oppositely to the case of
entanglement. This phenomenon reveals that quantum correlations
without entanglement have quite exotic characteristic from
entanglement. In addition, this effect also offers a solution to
make the entanglement of a system enhanced.

\begin{acknowledgments}
The work is supported in part by the NNSF of China Grant No.
90503009, No. 10775116, and 973 Program Grant No. 2005CB724508.
\end{acknowledgments}

\end{document}